\documentclass[a4paper, preprint,12pt,sort&compress,dvipdfmx]{elsarticle}

\usepackage{epsfig}
\usepackage{times} 
\usepackage{amssymb}
\usepackage{lineno}

\usepackage[utf8]{inputenc} 
\usepackage[T1]{fontenc} 
\usepackage{mathptmx} 

\journal{Journal of Magnetism and Magnetic Materials}

\begin{document}

\begin{frontmatter}

\title{Adiabatic temperature change in ErAl$_{2}$/metal PIT wires:
A practical method for estimating the magnetocaloric response of magnetocaloric composites}

\author[1]{Takafumi D. Yamamoto\corref{cor1}}
\ead{YAMAMOTO.Takafumi@nims.go.jp}
\author[1]{Hiroyuki Takeya}
\author[1]{Kensei Terashima}
\author[2]{Suguru Iwasaki}
\author[1,3]{Pedro Baptista de Castro}
\author[1]{Takenori Numazawa}
\author[1,3]{Yoshihiko Takano}

\cortext[cor1]{Correpsonding author}

\address[1]{National Institute for Materials Science, Tsukuba 305-0047, Japan}
\address[2]{Research Institute for Electronic Science, Hokkaido University, Sapporo 001-0020, Japan}
\address[3]{University of Tsukuba, Tsukuba 305-8577, Japan}

\begin{abstract}
We report the adiabatic temperature change in ErAl$_{2}$ magnetocaloric wires
fabricated by a powder-in-tube (PIT) process.
The adiabatic temperature change of the PIT wires is found to be determined
by not only the volume fraction of ErAl$_{2}$ core
but also the magnitude relationship between the specific heat of
the ErAl$_{2}$ core and the metal sheath.
We propose a quantitative analysis method for calculating
the temperature and core volume fraction dependence of
adiabatic temperature change in the PIT wire,
whose formula is applicable to also various magnetocaloric composites,
useful to estimate the magnetocaloric response prior to fabrication.
\end{abstract}

\begin{keyword}
Magnetocaloric effect \sep Analytical technique \sep Composites \sep Powder-in-tube method \sep Intermetallic compounds
\end{keyword}

\end{frontmatter}

\section{Introduction}
Magnetic refrigeration is a cooling technology
based on the magnetocaloric effect (MCE) of magnetic materials.
It can be better than conventional vapor-compression refrigeration
in that it is environmentally friendly, compressor-free,
and probably more efficient
\cite{Zimm-ACE-1998,Tegus-Nature-2002, Bruck-JPD-2005}.
Its potential applications range from
ultra-low temperature cryocoolers to room temperature refrigerators.
Recently, there has been increasing interest
in hydrogen liquefaction by magnetic refrigeration,
which is expected to play a key role in realizing
the so-called hydrogen society
\cite{Kamiya-Cryoc-2007,Utaki-Cryoc-2007,Matsumoto-JPCS-2009}.
With the aim of developing magnetic refrigeration applications,
the search for materials with large MCE
\cite{Gschneidner-RPP-2005,Franco-PMS-2018,Zhang-PhysicaB-2019}
and the development of effective refrigeration systems
\cite{Gschneidner-IJR-2008,Nielsen-IJR-2011,Numazawa-Cryo-2014}
have been actively studied.

The MCE is evaluated by
the magnetic entropy change ($\Delta S_{M}$)
and the adiabatic temperature change ($\Delta T_{\rm ad}$)
\cite{Tishin-Hand-1999}.
$\Delta S_{M}$ is the entropy change in a magnetic material
upon changing the magnetic field.
This quantity is related with the heat that the material exchanges with its surroundings,
and it determines the refrigerant capacity of a magnetic working substance
\cite{Wood-Cryogenics-1985}.
$\Delta T_{\rm ad}$ represents the change in temperature of a magnetic material
when magnetized (demagnetized) adiabatically,
which indirectly characterizes the temperature span
between the hot and the cold sides of a refrigerator
\cite{Gschneidner-AnnuRev-2000}.
The materials for the magnetic refrigeration should exhibit
large $\Delta S_{M}$ and $\Delta T_{\rm ad}$
in the working temperature range of each refrigeration device.

For practical use of candidate materials,
they are usually implemented in a refrigeration device as magnetic refrigerants
after being processed into various shapes such as plates and spheres
\cite{Kamiya-Cryoc-2007,Yu-IJR-2010,Tusek-IJR-2013}.
One way for shaping the candidate material is
to use an additional non-magnetic substance as a binder
\cite{Skokov-JAP-2014,Zhang-APL-2014,Zhang-JAP-2015,Liu-IEEE-2015,Krautz-Scripta-2015,
Pulko-JMMM-2015,Radulov-JMMM-2015,Radulov-IEEE-2015,Zhang-Scripta-2016,Matsumoto-JPConf-2017,
Radulov-IEEE-2017,Radulov-ActMater-2017,Wenkai-RMME-2017,Liu-Materials-2018,Si-SciRep-2018,
Wang-AEM-2018,Dong-JAC-2018,Zhong-JMMM-2018,Zhong-JMMM-2019},
where the resulting material is a kind of magnetocaloric composites.
On such an approach, one has to take care about
how the magnetocaloric properties of both $\Delta S_{M}$ and $\Delta T_{\rm ad}$ change
before and after the processing.
In this context, it would be time- and cost-saving
if the magnetocaloric performance of
the composite material could be predicted in advance.
Among $\Delta S_{M}$ and $\Delta T_{\rm ad}$,
$\Delta S_{M}$ is an extensive measure of the MCE
and is expected to mainly depend on
the volume fraction of magnetocaloric substances in the composite.
On the other hand, $\Delta T_{\rm ad}$ is an intensive measure of the MCE
and would be determined by more complex physical factors.
To our knowledge, there has been no study to analyze the effects of processing
on $\Delta T_{\rm ad}$ in magnetocaloric composites,
except for the pioneering work in Cu-covered magnetocaloric ribbons
\cite{Shao-JMMM-1996}.

A comparative study of several composites would give insights
into the adiabatic temperature change in magnetocaloric composites.
For this purpose, we focus on magnetocaloric wires fabricated by a PIT method,
in which a powder material filled in a metal tube is processed
into wired-shape by metal workings.
In our previous study \cite{TDY-JPD-2019},
we fabricated three types of PIT wires
consisting of ErAl$_{2}$ core material
and a non-magnetic metal sheath of Al, Cu, and Brass, respectively.
A cubic Laves phase ErAl$_{2}$ is
a candidate material for hydrogen magnetic refrigeration
with a Curie temperature $T_{\rm c}$ of $\sim$ 12 K
\cite{Nereson-JAP-1968, Hashimoto-ACM-1986}.
We have found that the magnetic entropy change per unit volume of PIT wire, $\Delta S_{M}^{\rm wire}$,
is reduced by decreasing the core volume fraction, regardless of the sheath materials.
This is a consequence of the extensive property of $\Delta S_{M}$.
We have further found that $\Delta S_{M}^{\rm wire}$ around $T_{\rm c}$ shows
a small difference depending on the sheath materials,
implying that the PIT process affects
the intrinsic magnetocaloric properties of the ErAl$_{2}$ core.

Here we present the investigation of the adiabatic temperature change
in the ErAl$_{2}$ PIT wires with different non-magnetic sheath materials.
We find a clear sheath material dependence on
the temperature dependence of $\Delta T_{\rm ad}$,
which is strikingly different from the situation in $\Delta S_{M}$
\cite{TDY-JPD-2019}.
We suggest a quantitative analysis method
and demonstrate that the magnitude relationship
between the specific heat of the constituent materials is
crucial for $\Delta T_{\rm ad}$ in the PIT wires,
in addition to the core volume fraction.
It is also discussed how to control $\Delta T_{\rm ad}$ with these two parameters.
\section{Experimental}
\begin{figure}[b]
\centering
\includegraphics[width=85.00mm]{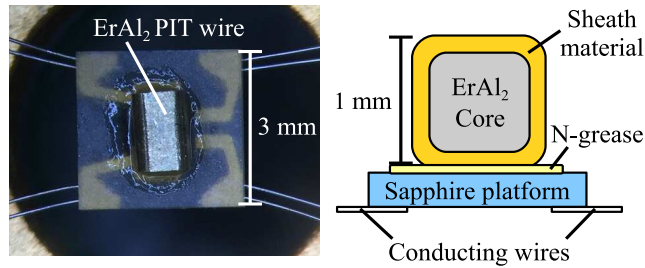}
\caption{(Color Online) Optical micrograph of a representative sample set-up
for the heat capacity measurements (the left) and
a schematic picture viewed from the side (the right).}
\label{fig:HCmeas}
\end{figure}

ErAl$_{2}$ PIT wires were fabricated by an \textit{ex-situ} PIT method
combined with a grooved rolling process.
Details of the preparation procedure were described elsewhere
\cite{TDY-JPD-2019}.
The resulting PIT wires were approximately 1 mm $\times$ 1 mm in size,
and the ErAl$_{2}$ core volume fraction ($\alpha$) was evaluated
to be about 0.49 for all the wires
through the magnetization measurements
\cite{TDY-JPD-2019}.
Bulk ErAl$_{2}$ was prepared as a reference sample by arc-melting,
followed by an annealing treatment in a vacuum at 1000 $^{\circ}$C for 24 h.
The X-ray diffraction measurement shows a bit trace (6\%) of
Er$_{2}$O$_{3}$ in the bulk ErAl$_{2}$.
Er$_{2}$O$_{3}$ is an antiferromagnet
with a N$\acute{\rm e}$el temperature of 3.4 K
\cite{Tang-PRB-1992},
but a small amount of this compound would not significantly affect
the physical properties of the ferromagnetic ErAl$_{2}$.

Heat capacity measurements were carried out by a thermal relaxation method
in the temperature ($T$) range between 2 and 60 K at zero magnetic field
($\mu_{0}H$ = 0 T) by using a Quantum Design PPMS.
As shown in Fig. \ref{fig:HCmeas},
each sample with a typical length of 1.4 mm was
put on the sapphire platform with N-grease.
Temperature dependence of entropy ($T$-$S$ curve) at 0 T was calculated from
that of the specific heat ($C$) at 0 T according to the formula
\begin{equation}
S(T,0) = \int_{0}^{T} \frac{C(T,0)}{T} dT.
\end{equation}
For each sample, the $T$-$S$ curves under various magnetic fields were obtained
by adding $\Delta S_{M} (T,\mu_{0} \Delta H)$
evaluated from the isofield magnetization measurements to $S(T,0)$,
where $\mu_{0} \Delta H$ represents the magnetic field change from 0 to $\mu_{0}H$.
In short, $S(T,\mu_{0}H) = S(T, 0) + \Delta S_{M} (T,\mu_{0} \Delta H)$.
Here the isofield magnetization was measured in zero-field cooling (ZFC) process.
After converting the unit of $S$ from J/g K to J/cm$^{3}$ K by multiplying the density,
the adiabatic temperature change was evaluated as
\begin{equation}
\Delta T_{\rm ad} (T,\mu_{0} \Delta H) = [T(S,\mu_{0} H) - T(S, 0)]_{S}
\end{equation}
from the calculated $T$-$S$ curves \cite{Pecharsky-JAP-1999,Pecharsky-JAP-2001}.
The density used here was the theoretical one for the bulk sample
and the one calculated from the measured mass and volume for each PIT wire.
Note that the same sample was used in the heat capacity and magnetization measurements.
%
\section{Results and discussion}
\begin{figure}[ht]
\centering
\includegraphics[width=65.00mm]{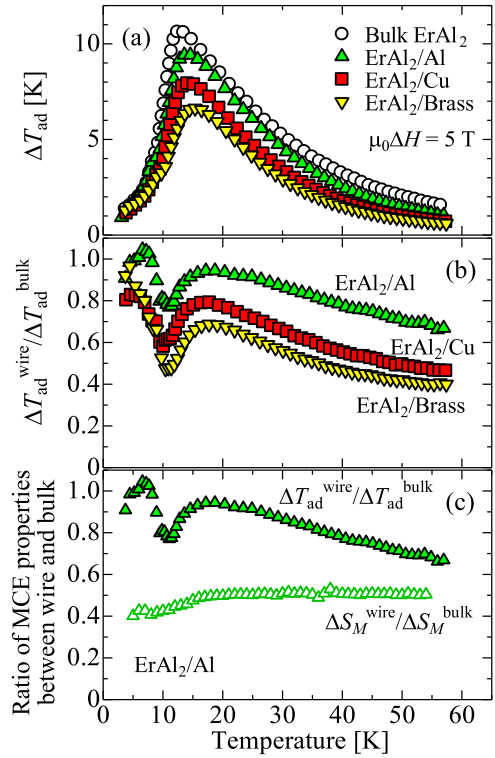}
\caption{(Color Online) (a) Temperature dependence of
the adiabatic temperature change for $\mu_{0} \Delta H$ = 5 T
in the ErAl$_{2}$ PIT wires ($\Delta T_{\rm ad}^{\rm wire}$)
and the bulk ErAl$_{2}$ ($\Delta T_{\rm ad}^{\rm bulk}$).
(b) The ratio of $\Delta T_{\rm ad}^{\rm wire}$ to $\Delta T_{\rm ad}^{\rm bulk}$
as a function of temperature for each wire.
(c) Comparison of $\Delta T_{\rm ad}^{\rm wire}$/$\Delta T_{\rm ad}^{\rm bulk}$
with $\Delta S_{M}^{\rm wire}$/$\Delta S_{M}^{\rm bulk}$ in the ErAl$_{2}$/Al wire.
The data of $\Delta S_{M}^{\rm wire}$/$\Delta S_{M}^{\rm bulk}$ is taken from
\cite{TDY-JPD-2019}.}
\label{fig:deltaT}
\end{figure}
Figure \ref{fig:deltaT}(a) shows the temperature dependence of
the adiabatic temperature change for $\mu_{0} \Delta H$ = 5 T
in the ErAl$_{2}$ PIT wires ($\Delta T_{\rm ad}^{\rm wire}$)
and the bulk ErAl$_{2}$ ($\Delta T_{\rm ad}^{\rm bulk}$).
Both $\Delta T_{\rm ad}^{\rm wire}$ and $\Delta T_{\rm ad}^{\rm bulk}$ exhibit
a broad peak structure with a maximum value at around $T_{\rm c}$ of 12 K,
but the magnitude of the former becomes smaller than that of the latter
as the sheath material changes in the order of Al, Cu, and Brass.
To make this sheath material dependence clear,
we plot the ratio of $\Delta T_{\rm ad}^{\rm wire}$ to
$\Delta T_{\rm ad}^{\rm bulk}$ as a function of temperature
for each PIT wire in Fig. \ref{fig:deltaT}(b).
The value of the ratio is about 0.9
at the lowest temperature for all the wires.
With increasing temperature,
$\Delta T_{\rm ad}^{\rm wire}$/$\Delta T_{\rm ad}^{\rm bulk}$ show
a dip structure around $T_{\rm c}$ and then gradually decreases at higher temperatures.
It is worth noting that
$\Delta T_{\rm ad}^{\rm wire}$/$\Delta T_{\rm ad}^{\rm bulk}$ differs depending on
the sheath material in the whole temperature range up to 60 K.
This is in sharp contrast to the fact that
$\Delta S_{M}^{\rm wire}$/$\Delta S_{M}^{\rm bulk}$ is
almost the same in all the PIT wires at above 30 K
\cite{TDY-JPD-2019}.
Even more surprisingly, as shown in Fig. \ref{fig:deltaT}(c),
$\Delta T_{\rm ad}^{\rm wire}$ of the ErAl$_{2}$/Al wire is only decreased by 20-30\%
compared to $\Delta T_{\rm ad}^{\rm bulk}$,
despite $\Delta S_{M}^{\rm wire}$ being reduced by half.
Besides, the temperature dependence is quite different between
$\Delta T_{\rm ad}^{\rm wire}$/$\Delta T_{\rm ad}^{\rm bulk}$
and $\Delta S_{M}^{\rm wire}$/$\Delta S_{M}^{\rm bulk}$,
as is evident in Fig. \ref{fig:deltaT}(c).
These three discrepancies between $\Delta T_{\rm ad}^{\rm wire}$ and $\Delta S_{M}^{\rm wire}$
indicate that $\Delta T_{\rm ad}^{\rm wire}$ is not likely to be simply determined by
the core volume fraction.
\begin{figure}[ht]
\centering
\includegraphics[width=70.00mm]{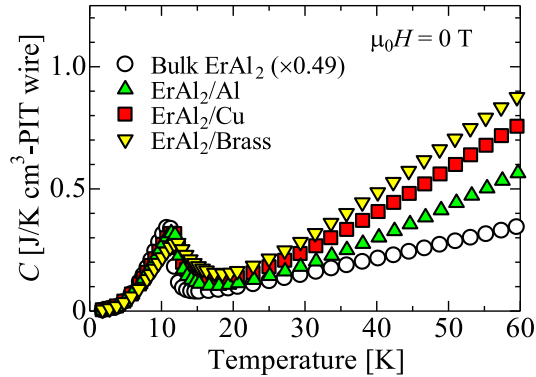}
\caption{(Color Online) Temperature dependence of the specific heat
at zero magnetic field for the ErAl$_{2}$ PIT wires and the bulk ErAl$_{2}$.
The data for the bulk sample is multiplied by the core volume fraction $\alpha$ of 0.49.}
\label{fig:C-0T}
\end{figure}

The adiabatic temperature change relates to
both the magnetic and thermal properties of a material,
\textit{i.e.}, the magnetization and the specific heat
\cite{Tishin-Hand-1999}.
To shed light on the nature of $\Delta T_{\rm ad}^{\rm wire}$,
we also examined the specific heat of the ErAl$_{2}$ PIT wires.
Figure \ref{fig:C-0T} shows the temperature dependence of the specific heat
at 0 T for the ErAl$_{2}$ PIT wires and the bulk ErAl$_{2}$,
where we take J per K per cubic centimeter of PIT wire as the unit of $C$.
The data for the bulk sample is multiplied by $\alpha$,
so that the corrected value can be regarded as the contribution of ErAl$_{2}$ core to $C$.
The specific heat of in all the PIT wires roughly coincide with
that of the bulk ErAl$_{2}$ at below 20 K,
in which a peak appears at around $T_{\rm c}$.
This fact implies that the ErAl$_{2}$ core dominates the specific heat of
the PIT wires in this temperature region.
At higher temperatures, $C$ differs significantly
between the PIT wires and the bulk sample.
Moreover, it strongly depends on the sheath material,
which is indicative of a substantial contribution of each sheath
on the total specific heat of the PIT wires.
From the above results, we find that
the specific heat in the PIT wires should be understood as
a summation of two components originating from
the ErAl$_{2}$ core and the sheath material.
Note that this is not the case in the magnetic properties
for which the contribution of non-magnetic sheath material can be negligible.

With this consideration in mind, let us discuss the adiabatic temperature change
in a PIT wire with a total volume of $V^{\rm wire}$,
which consists of a magnetocaloric core material
and a non-magnetic sheath.
When the PIT wire is magnetized (demagnetized) by a magnetic field change,
the core material produces the amount of heat
in response to its magnetic entropy change of $\Delta S_{M}^{\rm core}$.
The non-magnetic sheath does not act as a heat source, but rather as a heat load.
Accordingly, under an adiabatic condition,
the heat is used for not only the core material but also the sheath one,
resulting in the temperature change of $\Delta T_{\rm ad}^{\rm wire}$.
This can be described phenomenologically as
\begin{equation}
-T V^{\rm core} \Delta S_{M}^{\rm core}
= V^{\rm core} C^{\rm core} \Delta T_{\rm ad}^{\rm wire}
+ V^{\rm sheath} C^{\rm sheath} \Delta T_{\rm ad}^{\rm wire},
\label{eq:dissipation}
\end{equation}
where $V^{\rm core}$ and $C^{\rm core}$ are
the volume and the volumetric specific heat of the core material,
and $V^{\rm sheath}$ and $C^{\rm sheath}$ are those of the sheath material.
Note that $V^{\rm wire} = V^{\rm core} + V^{\rm sheath}$
and $C^{\rm core}$ and $C^{\rm sheath}$ are those under the final magnetic field.
Here we assume that the magnetic entropy change and the specific heat are the same
between the core material and its bulk counterpart.
Then, one can regard $-T \Delta S_{M}^{\rm core}/C^{\rm core}$ as $\Delta T_{\rm ad}^{\rm bulk}$.
Using that $\alpha = V^{\rm core}/V^{\rm wire}$,
we finally arrive at
\begin{equation}
\Delta T_{\rm ad}^{\rm wire} =
\Delta T_{\rm ad}^{\rm bulk}
\left(1+\frac{1-\alpha}{\alpha}\frac{C^{\rm sheath}}{C^{\rm core}} \right)^{-1}.
\label{eq:deltaT}
\end{equation}
We should note that in this model, the $\Delta S_{M}^{\rm wire}$ is determined solely by
the core volume fraction relative to the total volume of the PIT wire.
Namely, $\Delta S_{M}^{\rm wire} = \alpha \Delta S_{M}^{\rm core} = \alpha \Delta S_{M}^{\rm bulk}$.
By contrast, the core volume fraction dependence of $\Delta T_{\rm ad}^{\rm wire}$ is
not at all simple, as discussed below.

\begin{figure}[t]
\centering
\includegraphics[width=70.00mm]{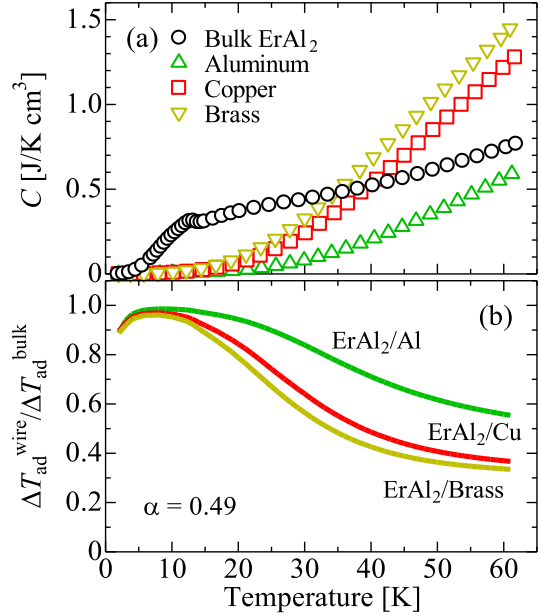}
\caption{(Color Online) (a)Temperature dependence of the volumetric specific heat at 5 T
in the bulk ErAl$_{2}$ and those at 0 T in the metals used as the sheath material.
(b) $\Delta T_{\rm ad}^{\rm wire}$/$\Delta T_{\rm ad}^{\rm bulk}$
calculated for each ErAl$_{2}$ PIT wire by using Eq. (\ref{eq:deltaT})
with $\alpha = 0.49$.}
\label{fig:Calculation}
\end{figure}
Figure \ref{fig:Calculation}(a) shows the temperature dependence of
the volumetric specific heat below 60 K at 5 T in the bulk ErAl$_{2}$
and those at 0 T in the metals used as a sheath material in this study.
We presume here that for the non-magnetic metals,
the specific heat at 0 T is the same as that at 5 T.
The specific heat of the bulk ErAl$_{2}$ is predominant
over those of sheath materials at temperatures below 20 K.
With increasing temperature,
the latter increase rapidly and become comparable to, or larger than the former.
These results indicate that $C^{\rm sheath}/C^{\rm core}$ in the PIT wires 
changes drastically up to 60 K, from near zero to 1 or more.
Based on Fig. \ref{fig:Calculation}(a),
we calculate $\Delta T_{\rm ad}^{\rm wire}$/$\Delta T_{\rm ad}^{\rm bulk}$
for each PIT wire by using Eq. (\ref{eq:deltaT}) with $\alpha$ = 0.49.
As shown in Fig. \ref{fig:Calculation}(b),
the calculated curves roughly reproduce the characteristics of
the experimental data in Fig. \ref{fig:deltaT}(b),
supporting the validity of the phenomenological model described by Eq. (\ref{eq:deltaT}).

However, we notice that no dip around $T_{\rm c}$ is observed in the calculated curves.
The dip in the experimental data probably originates from the intrinsic change
in $\Delta S_{M}$ of ErAl$_{2}$ core around $T_{\rm c}$ due to the PIT process
\cite{TDY-JPD-2019}.
Our model does not include this effect.
If there is no such intrinsic change, 
the experimental curves would become more resemble the calculated curves.
We further notice that $\Delta T_{\rm ad}^{\rm wire}$/$\Delta T_{\rm ad}^{\rm bulk}$ is
higher than 1 below 10 K for the ErAl$_{2}$/Al wire.
This is due to a small hump of $\Delta T_{\rm ad}$ around 8 K
seen in all the samples (not shown),
the size of which differs between the samples.
This hump is not observed for the data on the FC process,
so it is possibly attributed to some ZFC effects.
Such effects can be affected by some extrinsic factors, such as a residual magnetic field,
which may lead to the different size in hump depending on the sample.

Figure \ref{fig:Calculation} demonstrates that
not only $\alpha$ but also $C^{\rm sheath}/C^{\rm core}$ is an important parameter
to determine $\Delta T_{\rm ad}^{\rm wire}$/$\Delta T_{\rm ad}^{\rm bulk}$:
the smaller $C^{\rm sheath}/C^{\rm core}$,
the larger $\Delta T_{\rm ad}^{\rm wire}$/$\Delta T_{\rm ad}^{\rm bulk}$.
One finds that the specific heat is smaller for aluminum than bulk ErAl$_{2}$
over the entire measurement temperature range,
resulting in the largest $\Delta T_{\rm ad}^{\rm wire}$ observed in the ErAl$_{2}$/Al wire.
This should be because aluminum has the highest Debye temperature ($\Theta_{\rm D}$)
among the sheath materials
($\Theta_{\rm D}$ = 428, 343, and 319 K for Al, Cu, and Brass
\cite{Phillips-PR-1959,Veal-PR-1963}).
If the $C^{\rm sheath}$ is negligible, the heat is only used to
change the temperature of the core material, as is the case in its bulk form.
As a result, $\Delta T_{\rm ad}^{\rm wire}$ would be
comparable to $\Delta T_{\rm ad}^{\rm bulk}$.

\begin{figure}[t]
\centering
\includegraphics[width=85.00mm]{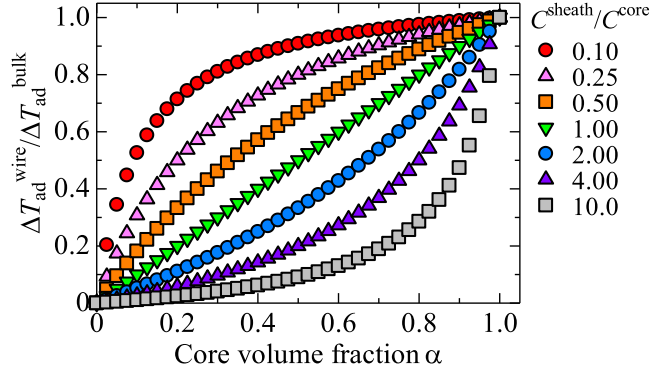}
\caption{(Color Online)
The core volume fraction dependence of
$\Delta T_{\rm ad}^{\rm wire}$/$\Delta T_{\rm ad}^{\rm bulk}$
for various values of $C^{\rm sheath}/C^{\rm core}$ ranging from 0.1 to 10.}
\label{fig:VFdep}
\end{figure}
In designing a magnetocaloric PIT wire,
it is useful to understand how the change in core volume fraction affects
the magnetocaloric performances in a PIT wire as compared to a bulk counterpart.
Figure \ref{fig:VFdep} shows the core volume fraction dependence of
$\Delta T_{\rm ad}^{\rm wire}$/$\Delta T_{\rm ad}^{\rm bulk}$
calculated based on Eq. (\ref{eq:deltaT})
for various values of $C^{\rm sheath}/C^{\rm core}$.
For $C^{\rm sheath}$ = $C^{\rm core}$,
$\Delta T_{\rm ad}^{\rm wire}$/$\Delta T_{\rm ad}^{\rm bulk}$ is proportional to $\alpha$,
as is $\Delta S_{M}^{\rm wire}$/$\Delta S_{M}^{\rm bulk}$.
On the other hand, the calculated curves bend downward when $C^{\rm sheath} > C^{\rm core}$,
by which the decrease in $\alpha$ sharply reduces
$\Delta T_{\rm ad}^{\rm wire}$/$\Delta T_{\rm ad}^{\rm bulk}$.
The most remarkable feature is found in the case for $C^{\rm sheath} < C^{\rm core}$,
where $\Delta T_{\rm ad}^{\rm wire}$/$\Delta T_{\rm ad}^{\rm bulk}$ becomes insensitive
to the decrease of $\alpha$.
Interestingly, $\Delta T_{\rm ad}^{\rm wire}$ is only 10\% less than
$\Delta T_{\rm ad}^{\rm bulk}$ for $C^{\rm sheath}/C^{\rm core} =$ 0.1,
even if $\alpha$ is decreased by 50\%.
As can be seen above,
the dependence of $\Delta T_{\rm ad}^{\rm wire}$/$\Delta T_{\rm ad}^{\rm bulk}$
on $\alpha$ can change dramatically with varying $C^{\rm sheath}/C^{\rm core}$.

Finally, we will point out some significant implications
from Figs. \ref{fig:Calculation} and \ref{fig:VFdep}. 
(i) In a PIT wire, an inevitable decrease in core volume fraction
may cause serious degradation of the magnetocaloric performances.
This is not always the case for $\Delta T_{\rm ad}$.
Under a certain condition, one can realize a large $\Delta T_{\rm ad}^{\rm wire}$
comparable to $\Delta T_{\rm ad}^{\rm bulk}$,
even though $\Delta S_{M}^{\rm wire}$ is largely reduced.
Such wires would be sufficient for use in active magnetic regenerator devices
\cite{Nielsen-IJR-2011,Barclay-Patent-1982,Smaili-Cryogenics-1998},
because these devices directly utilize $\Delta T_{\rm ad}$.
(ii) The influence of the core volume fraction and materials combination varies with
the transition temperature ($T_{\rm mag}$) of a core material.
Since the $C^{\rm sheath}$ generally becomes pronounced with increasing temperature,
the higher the $T_{\rm mag}$,
the greater the decline in $\Delta T_{\rm ad}^{\rm wire}$ at around $T_{\rm mag}$ is expected.
In such a case, one needs more core volume fraction.
(iii) On the other hand, for any $T_{\rm mag}$,
it is effective to select a sheath material
with a smaller $C^{\rm sheath}$ than $C^{\rm core}$.
At low temperatures below 100 K,
the volumetric specific heat is sensitive to the temperature,
and its value may vary significantly from material to material
according to the $\Theta_{\rm D}$.
In that sense, the advantage that $\Delta T_{\rm ad}^{\rm wire}$ is a function of
$C^{\rm sheath}/C^{\rm core}$ can be particularly exploited in this temperature range.
(iv) Shao et al. \cite{Shao-JMMM-1996} have investigated $\Delta T_{\rm ad}$ of
magnetocaloric ribbons consisting of Gd-based nanocomposite powder
and the surrounding Cu-sheath,
and suggested a quantitative analysis method similar to ours.
However, the formulas there are somewhat specific to the ribbons
in that the parameters are inherent to the ones.
It thus has not been fully examined how $\Delta T_{\rm ad}$ behaves
in magnetocaloric composites.
In the present study, we have illustrated that
the overall characteristics of $\Delta T_{\rm ad}$ in the PIT wire can be predicted
according to Eq. (\ref{eq:deltaT}).
This formula requires only the volume fraction of magnetocaloric substances
and the specific heat of the constituent materials.
Therefore, it would be more practical
and may be applied to various magnetocaloric composites
to estimate the adiabatic temperature change in advance.
%
\section{Summary}
We investigate the adiabatic temperature change of
ErAl$_{2}$ magnetocaloric PIT wires with different non-magnetic sheath materials.
Unlike the magnetic entropy change,
the characteristics of the adiabatic temperature change depend on
the sheath material over a wide temperature range.
We find that the magnitude relationship between the specific heat of
the constituent materials is a crucial parameter
for the adiabatic temperature change in the PIT wire,
and its temperature and core volume fraction dependence can be analyzed
using the formula based on the phenomenological model.
The analysis method proposed in this paper can be helpful for
designing the magnetocaloric composites
with relatively large adiabatic temperature change
comparable to the bulk counterpart.

\section*{Acknowledgements}
This work was supported by JST-Mirai Program Grant Number JPMJMI18A3, Japan.

\end{document}